# Impact of germline susceptibility variants in cancer genetic studies

Fernandez-de-Cossio, J.*†¥ and Perera, Y.‡¥

†Bioinformatics Department and ‡ Molecular Oncology Group, Direction of Biomedical Research, Center for Genetic Engineering and Biotechnology, Havana Cuba.
¥ Equal contribution
*Jorge.cossio@cigb.edu.cu

## Abstract

Although somatic mutations are the main contributor to cancer, underlying germline alterations may increase the risk of cancer, mold the somatic alteration landscape and cooperate with acquired mutations to promote the tumor onset and/or maintenance. Therefore, both tumor genome and germline sequence data have to be analyzed to have a more complete picture of the overall genetic foundation of the disease. To reinforce such notion we quantitatively assess the bias of restricting the analysis to somatic mutation data using mutational data from well-known cancer genes which displays both types of alterations, inherited and somatically acquired mutations.

## Introduction

In spite that germline susceptibility variants underlying advanced cancers generally evade to be clinically actionable, these inherited alterations are proving to be more relevant than anticipated (Ngeow & Eng, 2016). For instance, in a survey of 1566 patients nearly 16% of pathogenic variants linked to an inherited human disease were found in germline variants of 187 overlapping genes (Prasad et al., 2015). Moreover, using 466 DNAs from healthy tissues homozygous deletions (HD) totaling ~23 Mb and spanning 1% of protein-coding genes have been classified as germline HD (Bignell et al., 2010). Interestingly, loss of heterozygosity (LOH) for the wild-type allele of some tumor suppressor genes frequently associates with inherited germline mutations supporting the so called two hit hypothesis (Knudson, 1971; Nomoto et al., 2000; Saito et al., 2000). For instance, LOH for the wild-type allele have been found in 100% and 76% of cases with germline BRCA1 and BRCA2 truncations in ovarian cancer (Kanchi et al., 2014). Finally, cooperation between germline mutations and somatically acquired alterations within the same or different genes has been recently described in several tumor localizations (Lu et al., 2015). Altogether, these evidences implicate germline mutations/deletions as underlying active alterations modulating the risk to get cancer, influencing the tumor genome stability and cooperating with somatically acquired mutations to promote the cancer onset and/or maintenance.

Although somatic and germline genomes can be ascertained from studies that perform sequencing of tumor and normal-matched samples, most of the state-of-the-art multidimensional cancer genomics analysis tools (e.g. CBioPortal, UCSC Cancer Genome Browser) and driver gene prediction methods (e.g. IntoGen, MutSigCV) use filtered data; therefore restricting the analysis to somatically acquired mutations. While somatic alterations are the main contributor to cancer, the above mentioned evidences suggest inherited germline mutations imparts susceptibility and frequently shapes the acquired somatic alteration landscape (Lu et al., 2015). Therefore, both tumor genome and germline sequence data have to be analyzed to provide a more integral picture of the overall genetic contribution to disease (Kanchi et al., 2014). Here we quantitatively assess the impact of restricting the analysis to somatic mutation data in the study of the genetic contribution to cancer.

## Methods

We assume here that at least one of the variants considered in the study drives cancer development with detectable penetrance. When dealing with penetrance of genotypes epistasis and correlations are considered by construction. We disregard genotyping errors for the sake of simplicity.

Briefly, an individual with genotype $g$ has a probability $F_g$ to be drawn from a population (patient + non patient), and the chance (penetrance) an individual with genotype $g$ have cancer is denoted $p_g$. Consequently, the probability of drawing a cancer individual with genotype $c$ from the population is $p_c F_c$, and the probability of drawing a genotype $c$ individual from patients is $p_c F_c / \sum_g p_g F_g$, where the denominator is the probability of drawing a cancer individual from the population, and $g$ run within the possible genotypes.

The expected number $n_c$ of individual with genotype $c$ in a patient sample of size $n$ is

$$\langle n_c \rangle \sim n \frac{p_c F_c}{\sum_g p_g F_g}$$

The penetrance $p_c$ of this alteration can be approximate by

$$\langle p_c \rangle \sim \frac{n_c}{n F_c} \sum_g p_g F_g$$

Let $\alpha_g > 0$ and $1 - \alpha_g$ be the relative proportions of somatic and germinal alteration of a particular genotype variant $g$ found in cancer individuals. I.e. $\alpha_c \langle n_c \rangle$ patients are expected to acquire the alteration $c$ somatically. The expected number of patients $n'_c$ with somatic mutation $c$ is $\langle n'_c \rangle = \alpha_c \langle n_c \rangle$. When non-somatic alteration are filtered out from the data, penetrance of gene $c$ is underestimated in the same proportion

$$p'_c = \alpha_c p_c$$

**Single locus**: Let $A$ and $a$ be normal and malign allele's variants at locus $A$. Let $P_A$ and $P_a = 1 - P_A$ be the probability that the normal and malign allele came up from the germline.

**Interactions**: Let consider two interacting locus A and B with normal and malign variant allele's A|a and B|b respectively, and $\alpha$ and $\beta$ be each the relative proportions of somatic and germinal alterations in either locus respectively. When both malign alteration $a$ and $b$ co-occur in a patient, the chance this interacting variant a-b is counted by the somatic filtering is

$$P_A \alpha P_B \beta,$$

that is the chance both malignant alterations were somatically acquired. The chance this interacting is undercounted is

$$P_a(1-\alpha)P_B\beta + P_A\alpha P_b(1-\beta) + P_a(1-\alpha)P_b(1-\beta),$$

that is the chance that at least one of the malignant variant being not somatically acquired, which equal the chance tumor variants $a$ and $b$ (acquired or not somatically) co-occur, minus the chance both were acquired somatically.

$$[P_a(1-\alpha) + P_A\alpha][P_b(1-\beta) + P_B\beta] - P_A\alpha P_B\beta$$

Non-malign variants A and B can co-occur somatically by chance with probability $P_a\alpha P_b\beta$ in a patient sample; hence they are reported by the filtering process. The competence of this false interaction hit is unfair in the filtering scenario which undercounts the actual a-b interactions.

The issue is magnified for those tumors requiring the cooperation between more than two driver alterations. For example, the chance $n$-order interaction of driver genes is undercounted is

$$\prod_{i \in \text{loci}}^{n}[P_{a_i}(1-\alpha_i) + P_{A_i}\alpha_i] - \prod_{i \in \text{loci}}^{n} P_{A_i}\alpha_i$$

Where $A_i|a_i$ are the normal and malign variant allele's of locus $i$ and $\alpha_i$ is the relative proportions of somatic alteration in locus $i$, for $n$ interacting locus.

## Results and Discussion

The filtering process leading to available somatic mutation data from TCGA confined to SNV for easy illustration is sketched in Table 1. In this hypothetical data sample suppose that variant $C$ is a malignant predisposing alteration in a given position of gene1. Since this variant appear somatically only in partient1, the same malign variant is not counted for patient2, which is carried un-mutated from the germline.

*Table 1: A sketch of the filtering process leading to available somatic data. The symbol ">" indicates the occurrence of a somatic mutation in a given gene position (orange background). For example G > C denotes a germline variant G mutated somatically to variant C in a given position of gene1 in patient 1. A single symbol (nucleotide) represent un-mutated germline variant, which are not reported (while background).*

|          | gene1 | gene2 | gene3 | gene4 | gene5 | gene5 |
|----------|-------|-------|-------|-------|-------|-------|
| patient1 | G > C | A > T | A > G | A     | T > G | T     |
| patient2 | C     | A > T | G     | A > C | T     | A     |

In general, those positions of a genome which are not somatically mutated are undercounted. These omission might have not relevant implications provided the relative frequency of the malign variant in the population can be dismissed with respect to those appearing somatically (i.e. $\alpha \sim 1$, as would be the commonest cases), ex. tumor suppressor gene TP53 exhibit a comparatively larger rate of somatic mutations (Kanchi et al., 2014). However, inherited alteration with relatively small somatic mutation $\alpha$, rarely differ from the matched germinal line and are typically filtered out. For example malign variants of BRCA1 and BRCA2 genes are twice more frequently inherited than somatically acquired (Kanchi et al., 2014). These malign variant found recurrently in breast and ovarian cancer germlines, are undoubtedly undercounted in the filtered data of somatic mutations, and current renowned software tools reflect the issue.

The somatic filtering (undercounting) occur with probability $P_a(1-\alpha)$, that is the chance of having the malign allele $a$, being it not somatically acquired. On the other hand, non-malign variant A can co-occur somatically with probability $P_a\alpha$ in individuals, hence they are reported by the filtering process. This false hit can introduce a nuisance competence in the filtering scenario which undercount the actual $a$ variant. The undercounting of single alteration is detailed in Table 2.

*Table 2*

| alterations | probability | True count | Actual count | True value | Bias |
|---|---|---|---|---|---|
| A > a | $P_A \alpha$ | +1 | +1 | TP | No |
| a | $P_a(1-\alpha)$ | +1 | 0 | FN | Yes |
| a > A | $P_a \alpha$ | +1 | +1 | FP | Yes |
| A | $P_A(1-\alpha)$ | 0 | 0 | TN | No |
| expected | | $\alpha + P_a(1-\alpha)$ | $\alpha$ | | |

For example, a recent study of ovarian cancer (Supplementary Table 3, in Kanchi et al., 2014), BRCA1 alterations were found in 71 of 429 cases, 50 of them were classified as germline alterations, and 21 as somatic mutations. By draft estimation we assign

$$P_a = \frac{21+50}{429}, \quad \alpha = \frac{21}{21+50},$$

The relative bias

$$\frac{P_a(1-\alpha)}{\alpha + P_a(1-\alpha)} = 0.282666$$

Relative bias indicated that about 28% on average of the actual alteration hits are lost by the somatic filtering procedure. Similarly, BRCA2 alterations were found in 47 of 429 cases, 36 of them were classified as germline alterations, and 11 as somatic mutations. The same estimation above yields that 26% of the malignant alterations hits are filtered out.

The way somatic mutations data is made available imply that the set of locus whose variants are reported for one individual are not the same set of locus reported for other individual. The variant of a missing locus can then be implicitly regarded as a variant different to those alterations found somatically in the sample, since the count is not increased by neither of them.

Let us consider the extremal case of an alteration $c$ with high penetrance and very low rate of somatic incidence i.e. $\alpha \sim 0$. If only somatic mutation are considered from the data, the chance to report this predominantly hereditary alteration in patients samples becomes very low $\langle n'_c \rangle \sim 0$, even if $\langle n_c \rangle$ were not small. In the most lucky case when $0 \ll n'_c$, the penetrance estimation is underestimated, and association with driving mechanism are mostly biased. Further, as this high penetrance mutation passed unnoticed (false negative), even when it probably was the single alteration driving cell to cancer, the guilt is forcedly spread to other somatic alterations with sufficient counting statistic to cover the missed cause (false positive). This is not an unrealistic case, considering that 39 of 592 alterations reported in the cancer gene census are inherit alterations. Things can turned worse when gene interactions are examined. Let consider two locus A and B with variant allele's a|A and b|B respectively. Population frequencies and penetrance are as follow:

Let $\alpha$ and $\beta$ be the proportion of somatic mutation in locus A and B respectively.

| germline variants pair | Tumor variants pairs | | | |
|---|---|---|---|---|
| | ab | aB | Ab | AB |
| ab | $P_a(1-\alpha)P_b(1-\beta)$ | $P_a(1-\alpha)P_b\beta$ | $P_a\alpha P_b(1-\beta)$ | $P_a\alpha P_b\beta$ |
| aB | $P_a(1-\alpha)P_B\beta$ | $P_a(1-\alpha)P_B(1-\beta)$ | $P_a\alpha P_B\beta$ | $P_a(1-\alpha)P_b\beta$ |
| Ab | $P_A\alpha P_b(1-\beta)$ | $P_A\alpha P_b\beta$ | $P_A(\alpha-1)P_b(1-\beta)$ | $P_A(1-\alpha)P_b\beta$ |
| AB | $P_A\alpha P_B\beta$ | $P_A\alpha P_B(1-\beta)$ | $P_A(1-\alpha)P_B\beta$ | $P_A(1-\alpha)P_B(1-\beta)$ |

| germline variants pair | Tumor variants pairs | | | |
|---|---|---|---|---|
| | ab | aB | Ab | AB |
| ab | 0 | 0 | 0 | 1 |
| aB | 0 | 0 | 1 | 0 |
| Ab | 0 | 1 | 0 | 0 |
| AB | 1 | 0 | 0 | 0 |

The interaction undercounting bias occur with probability
$$P_a(1-\alpha)P_b(1-\beta) + P_a(1-\alpha)P_B\beta + P_A\alpha P_b(1-\beta)$$

Importantly, inherited alterations with small somatic mutation rate rarely differ from the matched germinal line sample and are typically filtered out. Highly penetrant driver alterations can pass unnoticed by this procedure. Furthermore, significant interactions can be missed (false negative) and those counting statistics can amount for association between innocuous variant pairs (false positive interactions).

The use of unfiltered germline data implies the analysis of huge volumes of information and its handling require granted access to controlled data (Jones et al., 2015). However, ignoring germline mutations underestimates the actual frequency of pathogenic alterations, and consequently overestimates non-pathogenic frequency in patient samples. Furthermore, the filtering bias might impact the prediction of cancer drivers and mutually exclusive alterations, as well as the survival analysis performed by current analysis tools.

For instance, IntOGen collects and analyses somatic mutations in thousands of tumor genomes to identify cancer driver genes by mutation frequency and signals of positive selection on protein-coding genes (Gonzalez-perez et al., 2013). However, by looking only in tumor somatic data the tool may be ignoring roughly twice the BRCA1/BRCA2 mutations that are actually reported (Kanchi et al., 2014). This bias may also significantly impact at least two major outputs from cBioPortal (Gao et al., 2014). Mutually Exclusive alterations and Survival analysis may be influenced when genes which are mutated at both germline and somatic level are included in the analysis. Of note, the Cancer Gene Census comprises roughly 97 of 595 genes with germline mutations and 58 of them mutated at both levels (Futreal et al., 2004). Altogether, biased analysis are expected to be related to the frequency of rare pathogenic germline truncations underlying each tumor localization, from a mild anticipated impact on AML (roughly 4% of pathogenic germline truncations) to a potentially broader effect in ovarian cancer analysis (19%) (Lu et al., 2015). Remarkable, recent woks start to take into consideration full mutational patient data to analyze clinical significance of individual gene alterations and the potential interactions derived thereof (Kanchi et al., 2014; Lu et al., 2015).

## Conclusion

Malignant inherited alterations circulating in the population can be elusive to detection, analysis and clinical interpretation when genomic data is limited to filtered somatic alterations. Furthermore, significant interactions between inherited and somatically acquired alterations may also being overlooked in cancer patients. Here, we estimated the bias introduced in common cancer genome analysis when mutational data is restricted to somatic alterations, and illustrated the potential implications of ignoring germline sequence data from normal-matched samples with well-known examples. Taking into account the growing list of identified pathogenic germline variants, tumor genome and germline sequence data is required to approach a more complete picture of the overall genetic foundation of the disease.